# A CRITICAL IMPROVEMENT ON OPEN SHOP SCHEDULING ALGORITHM FOR ROUTING IN INTERCONNECTION NETWORKS

Stavros Birmpilis and Timotheos Aslanidis

National Technical University of Athens, Athens, Greece

## ABSTRACT

*In the past years, Interconnection Networks have been used quite often and especially in applications where parallelization is critical. Message packets transmitted through such networks can be interrupted using buffers in order to maximize network usage and minimize the time required for all messages to reach their destination. However, preempting a packet will result in topology reconfiguration and consequently in time cost. The problem of scheduling message packets through such a network is referred to as PBS and is known to be NP-Hard. In this paper we haveimproved, critically, variations of polynomially solvable instances of Open Shop to approximate PBS. We have combined these variations and called the induced algorithmI_HSA (Improved Hybridic Scheduling Algorithm). We ran experiments to establish the efficiency of I_HSA and found that in all datasets used it produces schedules very close to the optimal. In addition, we tested I_HSA with datasets that follow non-uniform distributions and provided statistical data which illustrates better its performance.To further establish I_HSA's efficiency we ran tests to compare it to SGA, another algorithm which when tested in the past has yielded excellent results.*

## 1. PROBLEM DESCRIPTION- INTERCONNECTION NETWORKS

In the past years,Interconnection Networks have been used quite often and in many different topologies. In this paper, we consider Multistage Interconnection Networks, which they provide high speed services with a large bandwidth and are therefore ideal in providing quality communication services in supercomputers, routers, clusters of machines and generally where there is a need for parallelization. In a Multistage Interconnection Network, switches are installed between the source and destination nodes, and thus, the topology is defined dynamically. Of course, the number of switches is much less than the number of possible paths, because otherwise it would be too expensive to construct. Therefore, there are always paths which are blocked, and others which are open for transmission. Message packets can be interrupted using buffers in order to maximize network usage and minimize the time required for all packets to reach their destination. However, pre-empting a packet will result in time cost, as the networks switches will need to reconfigure. Time to setup for the next packet can often be significant. The problem of scheduling message packets through such a network is referred to as PBS (Preemptive Bipartite Scheduling). Even though numerous algorithms have been designed in an effort to produce efficient schedules there seems to still exist room for further research.

## 2. THE GRAPH MODEL FOR PBS

As there are 2 sets between which the multiplexing and demultiplexing process takes place the ideal representation seems to be a bipartite graph $G(V, U, E, w)$. Source stations will be assigned to $V$, destination stations to $U$, messages to be transmitted will be the edges connecting nodes of $V$ to nodes of $U$. $w : E \rightarrow \mathbb{Q}_+$ will be a weight function giving each edge $e = (v, u)$ a weight equal to the duration of the transmission for $v$ to $u$. Given a matching $M$ in $G$ we will denote by $w(M)$ the maximum weight of any edge $e \in M$, that is $w(M) = max\{w(e), e \in M\}$. Following the notation used in previous research on the problem, $\Delta$ will denote the degree of $G$, $W$ the maximum sum of edge weights incident to any of the nodes and $d$ the setup cost to prepare for the next packet transmission. Thus, a feasible schedule for PBS would cost $\sum_{i=1}^{N} w(M_i) + d \cdot N$, where $N$ is the number of times the network has to reconfigure so that all data will be transferred.

Using these notations, the value $L = W + d \cdot \Delta$ represents a lower bound. $L$ is not always achievable but is easy to calculate and is considered to be a good approximation of the optimal solution when designing near optimal algorithms for PBS.

## 3. PAST RESEARCH ON PBS

The NP-Hardness of PBS derives from the fact that it is a bicriteria minimization problem, namely the objective function to be minimized depends on two different criteria each of which affects the other. Regardless the hardness of minimizing both criteria simultaneously, minimization of each criterion separately is relatively easy. Algorithms proposed by the authors of [11] and [9] minimize the number of preemptions while the one in [16] minimizes the transmission time. In general, the problem is $4/3 - \varepsilon$ inapproximable for all $\varepsilon > 0$ as shown in [7]. The best guaranteed approximation ratio of any algorithm proposed for the problem is $2 - \frac{1}{d+1}$. Proof of that can be found in [1]. Many other algorithms have been proposed in order to provide solutions close to the optimal. Experimenting on test cases has yielded good results in [3], [4] and [8].

In this manuscript we try to exploit in the best way possible a reduction of PBS to the open shop scheduling problem ($Om| |C_{max}$), in order to use polynomial time algorithms proposed for some special instances of it, to minimize each criterion separately and combine the results to design an improved hybrid algorithm (I_HSA – Improved Hybridic Scheduling Algorithm), which will tackle the bicriteria problem efficiently.

## 4. REDUCING PBS TO OPEN SHOP AND DESIGNING IMPROVED HSA

*Theorem 1:* Any instance of PBS can be transformed to an instance of Open Shop and vice versa.
*Proof:* Let $G(V, U, E, w)$ be the graph corresponding a PBS instance. We transform this graph to an open shop instance in the following way: $V = \{v_1, v_2, ..., v_n\}$ will be the set of processors $P = \{P_1, P_2, ..., P_n\}$, $U = \{u_1, u_2, ..., u_m\}$ will be the set of Jobs $J = \{J_1, J_2, ..., J_m\}$ and $E = \{(v_i, u_k)|v_i \in V, u_k \in U\}$ will be the set of operations $O = \{O_{ik}| i = 1, 2, ..., n \text{ and } k = 1, 2, ..., m\}$. $O_{ik}$ is the opeari on of $J_k$ to be processed on processor $P_i$. The processing time of each operation will be calculated by the function $p : O \rightarrow \mathbb{Q}_+$, where $p(O_{ik}) = \begin{cases} w(v_i, u_k), if \ (v_i, u_k) \in E \\ 0, otherwise \end{cases}$.

The inverse transformation is straight forward.

Unfortunately, the above reduction does not imply of a way to solve PBS using open shop algorithms as a PBS schedule would preempt all transmission simultaneously, while open shop scheduling does not have such a requirement. Yet, there exist two special instances of the Open Shop problem that are known to be solvable in polynomial time and are exactly right for our purposes. $Om|prmp|C_{max}$ in which preemption is allowed and $Om|p_{ij} = 0,1|C_{max}$ in which all processing times are either 0 or 1.

The polynomial time algorithm described in [16] minimizes a preemptive open shop makespan by preempting all processor tasks simultaneously. We will refer to this algorithm by LLA (Lawler-Labetoulle Algorithm). LLA uses linear programming techniques to define a set of tasks in order to reduce the workload of all stations that, in each step of the process are assigned with the maximum workload $W$. The authors of [16] call this a decrementing set. The number of preemptions is $O(m^2 + n^2)$. In order to improve the results of LLA instead of using a random decrementing set to reduce the workload of the stations we use one produced by a maximum weighted perfect matching algorithm. We will call this variation of LLA, POSA (Preemptive Open Shop Algorithm). We will use POSA to minimize I_HSA's makespan.

To complete I_HSA we also need an algorithm which will minimize the number of preemptions. A linear programming algorithm for $Om|p_{ij} = 0,1|C_{max}$ is described in [9]. Yet, in order to better fit the requirements of network transmission, the authors in [2] used an Open Shop algorithm which they called OS01PT.

*OS01PT Algorithm (Open Shop 0, 1 Processing Times)*
Step1: Add the minimum number of nodes needed to $G(V,U,E)$ so that $|U| = |V|$. Call the induced graph $G'$
Step2: Add edges to $G'$ to make it degree-regular.
Step3: Assign weights to the edges of $G'$ in the following way: Edges of the initial graph will weigh 1, while newly added edges in Step2 will weigh 0.
Step4: Calculate a perfect matching $M$ in $G'$.
Step5: Remove all edges of $M$ from $G'$.
Step6: Repeat Step4 and Step5 until $G' = \emptyset$.

To make the graph degree regular they used the subroutine described in [11].

*Theorem:* OS01PT will produce a schedule for PBS with exactly $\Delta$ transmissions.
*Proof:* By induction on the value of $\Delta$.

For $\Delta = 1$: Since $G'$ is degree regular, all nodes have exactly one adjacent edge. These edges form a perfect matching for $G'$ and the transmission will conclude in one step.

Let the theorem stand for any regular graph with $\Delta = n - 1$.

Suppose that $\Delta = n$. A perfect matching in $G'$ will reduce the degree of all nodes by one, thus making $G'$'s degree $n - 1$. From the inductive hypothesis an $n - 1$ degree graph will need $n - 1$ transmissions to schedule its data. Therefore, to transmit all data $1 + (n - 1) = n$ transmissions will be needed. Proof that a perfect matching can always be found in a graph with $\Delta = n > 1$ can also be found in [11].

In order for OS01PT to better suit the requirements of the problem in [2], instead of calculating a perfect matching as described in Step4 of the algorithm's description, the authors used a maximum weighted perfect matching algorithm just as in the case of POSA.

Well, in the same direction we took it one step further and we designed an improvement for this algorithm, which produces much better results.

*I_OS01PT Algorithm (Improved - Open Shop 0, 1 Processing Times)*

Step1:  Add the minimum number of nodes needed to $G(V, U, E)$ so that $|U| = |V|$. Call the induced graph $G'$

Step2:  Add edges to $G'$ to make it degree-regular.

Step3:  Assign weights to the edges of $G'$ in the following way: Edges of the initial graph will weigh $w' = w + d$ where $w$ is the initial weight and $d$ the setup cost, while newly added edges in Step2 will weigh 0.

Step4:  Calculate a maximum weighted matching $M$ in $G'$.

Step5:  Remove all edges of $M$ from $G'$.

Step6:  Repeat Step4 and Step5 until $G' = \emptyset$.

This heuristic does not always achieve the minimum number of preemptions but it appears to perform much better on average.

The reason that this algorithm gives better results in our problem, as we can see in the graph below, is that as $d$ is increasing, the same happens for the need to reduce the degree of the graph, otherwise it will cost more. With this variable in the algorithm, we set a priority for the matching algorithm in Step4 towards the edges which, when removed, will reduce the degree of the graph.

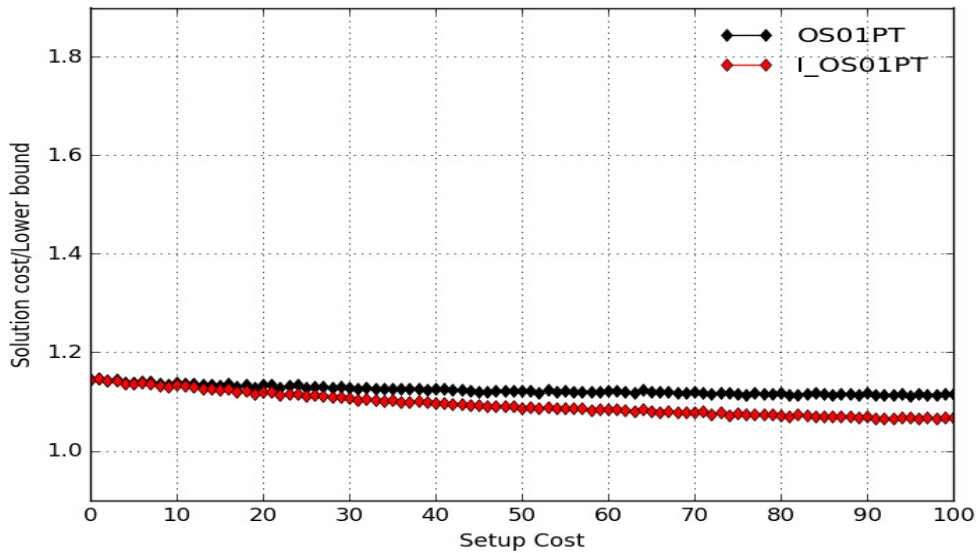

We now have all the necessary tools to design Improved HSA:

*I_HSA (Improved Hybridic Scheduling Algorithm)*

Step1:  Let $S_1$ be the feasible schedule produced for PBS using POSA. Let $C_1$ be the cost of $S_1$.

Step2: Let $S_2$ be the feasible schedule produced for PBS using I_OS01PT. Let $C_2$ be the cost of $S_2$.

Step3: If $C_2 < C_1$ then transmit as in $S_2$, else transmit as in $S_1$.

# 5. COMPUTATIONAL COMPLEXITY OF IMPROVED HSA

First of all, we suppose for simplicity that $|V| = |U| = n$.

Now, it is true that for producing a single schedule for the network, the computational complexity is the same for both of our algorithms (POSA and I_OS01PT). And it is: $S(n) = O(n^4) + \Theta(n)$. However, the number of single schedules, that will comprise a full schedule, differentiate in each case of the I_HSA. When using POSA, this number will be $O(n \cdot p_{max})$, where $p_{max}$ is the maximum possible weight (or message duration), and when using I_OS01PT, it will be $O(n)$.

Therefore, $I\_HSA(n, d) = \begin{cases} O(n \cdot p_{max}) \cdot S(n), & if\ d < d_{critical} \\ O(n) \cdot S(n), & if\ d > d_{critical} \end{cases}$

# 6. DECIDING A CRITICAL VALUE OF d FOR IMPROVED HSA

Five hundred test cases have been run for a 30 source-30 destination system for values of setup cost varying from 0 to 100 and message durations varying from 0 to 120. We have to point out that since PBS is an NP-Hard problem, calculating an optimal schedule is inefficient therefore to estimate the approximation ratio we have used the lower bound to the optimal solution namely $W + \Delta \cdot d$.

Figure 1 depicts the deviation from the optimal solution when using POSA to calculate a schedule for PBS. Figure 2 depicts the corresponding results when using I_OS01PT, while Figure3 shows the results yielded by I_HSA. The test cases in these figures are following a uniform distribution.

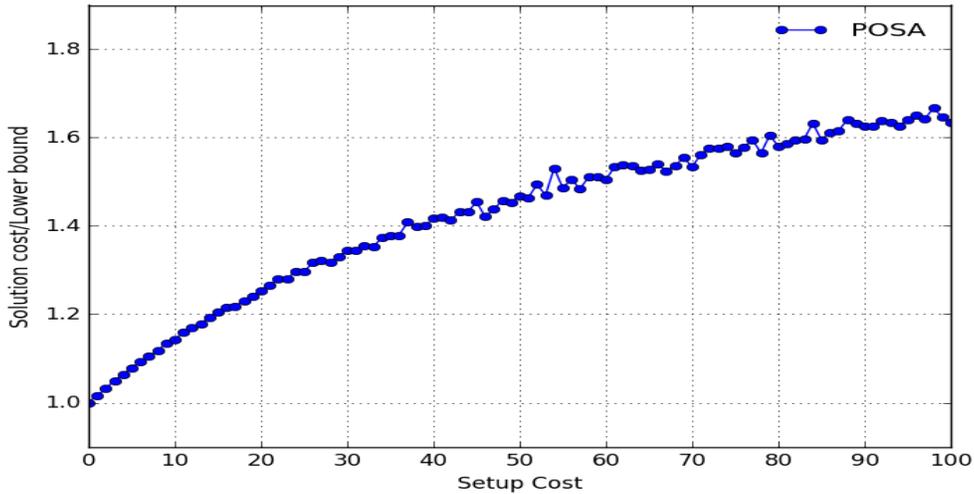

Figure 1. Average Solution cost/lower bound using POSA

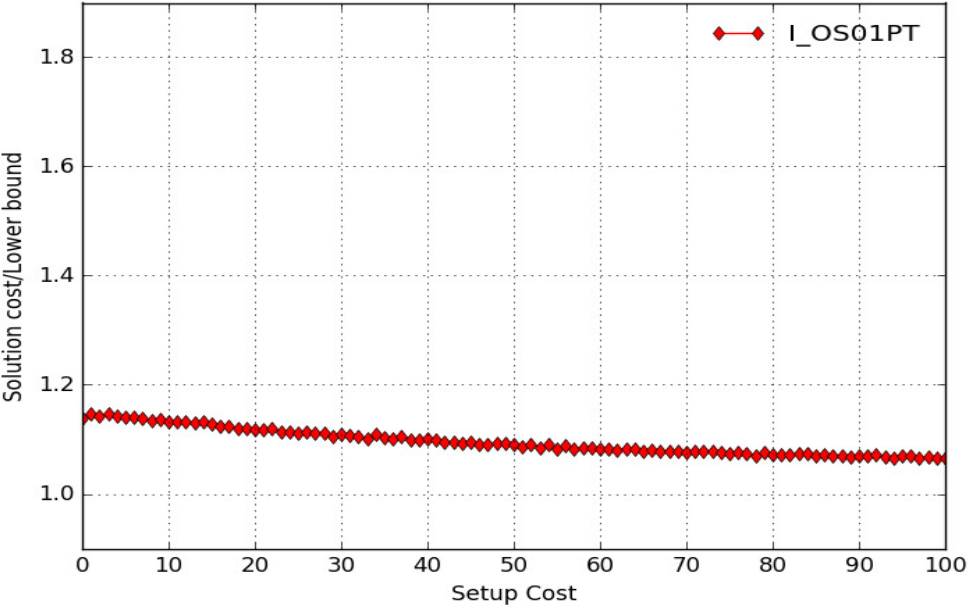

Figure 2. Average Solution cost/lower bound using I_OS01PT

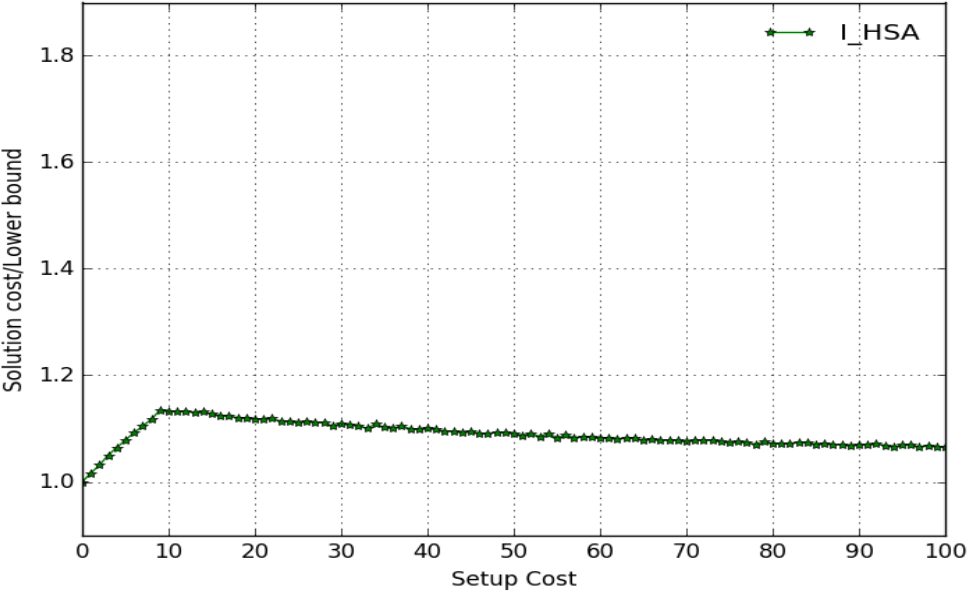

Figure 3. Average Solution cost/lower bound using I_HSA (uniform)

According to Figure1 and Figure 2, the appropriate value of $d$ to switch from POSA to I_OS01PT is $d = 9$.

Figure 4 shows the $(worst\ solution\ cost)/(lower\ bound)$ ratio of I_HSA for any of the instances used for each value of $d$. Note that it never exceeds 1.3.

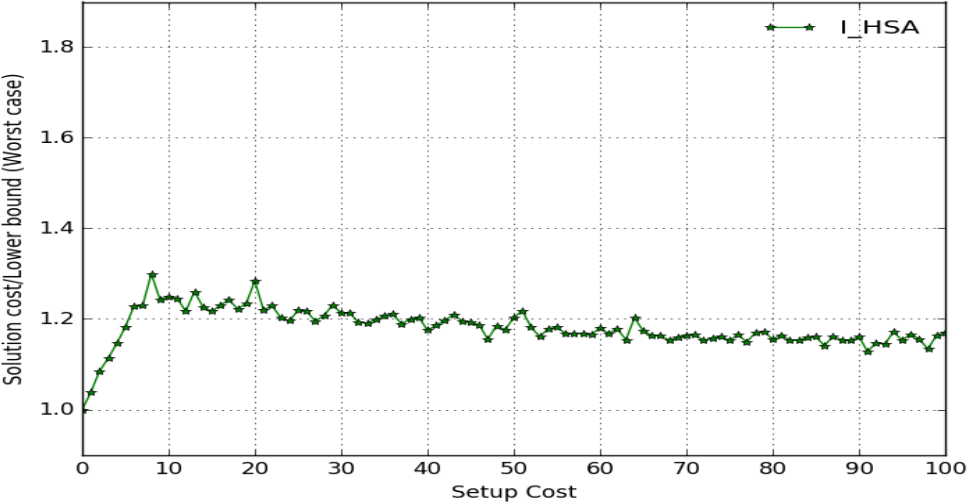

Figure 4. Worst solution cost/lower bound using I_HSA.

Moreover, Figure 5 shows the results yielded by running I_HSA on test cases following normal distribution and Figure 6 following exponential distribution. The critical value of d in the caseof normal distribution is at $d = 8$ and of exponential distribution is at $d = 16$.

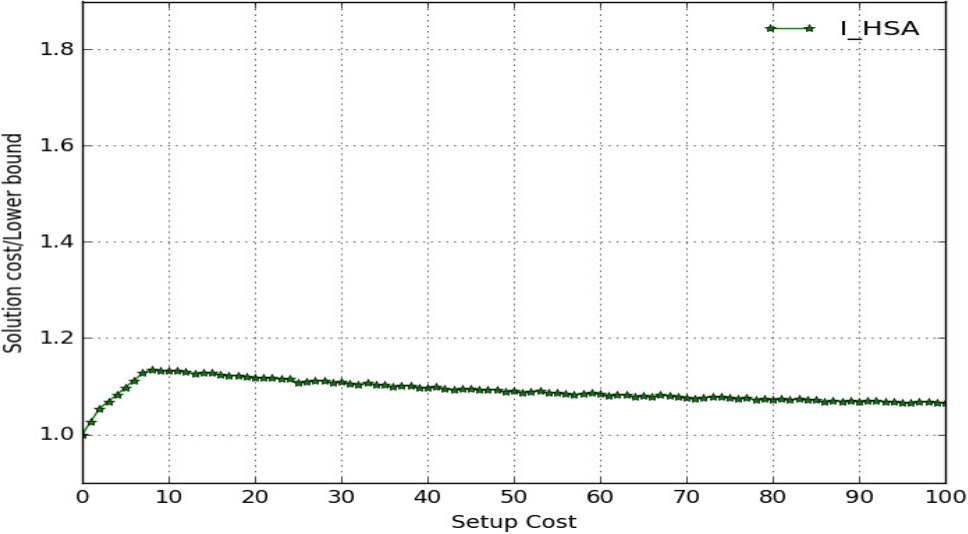

Figure 5. Average Solution cost/lower bound using I_HSA (normal)

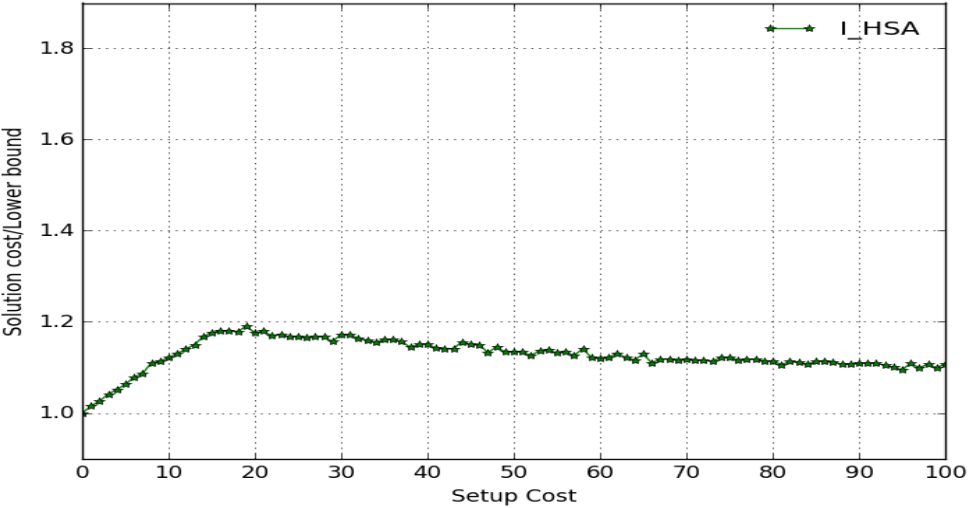

Figure 6. Average Solution cost/lower bound using I_HSA (exponential)

# 7. ILLUSTRATIVE STATISTICS ON THE RESULTS OF THE VARIATIONS CONCERNING THE OTHER CRITERION

On the tables below, we can see that the algorithms POSA and I_OS01PT are probably the most suitable for our problem. That is because, even though, they are designed to minimize only the one out of the two criteria, they still manage to tackle the other criterion efficiently as well.

| $d$ | POSA $-N/\Delta$ |
| --- | --- |
| 0 | 1.09375 |
| 10 | 1.06055 |
| 20 | 1.03516 |
| 30 | 1.02344 |
| 40 | 1.01367 |
| 50 | 1.01366 |
| 60 | 1.01563 |
| 70 | 1.01953 |
| 80 | 1.00977 |
| 90 | 1.01172 |
| 100 | 1.02148 |

| $d$ | I_OS01PT $- \sum_{i=1}^{N} w(M_i)/W$ |
|-----|------------------------------------|
| 0 | 1.14774 |
| 10 | 1.15503 |
| 20 | 1.15735 |
| 30 | 1.14184 |
| 40 | 1.15571 |
| 50 | 1.15190 |
| 60 | 1.14587 |
| 70 | 1.14580 |
| 80 | 1.13574 |
| 90 | 1.14801 |
| 100 | 1.14980 |

## 8. COMPARING IMPROVED HSA TO ANOTHER EFFICIENT ALGORITHM FOR PBS

One of the most efficient algorithms designed by researchers for PBS in the past is the SGA (Split Graph Algorithm). SGA splits the initial graph in two subgraphs, one with messages of duration less than d and another one with messages of duration at least d. The larger messages arescheduled first and then the small ones. It was found to be very efficient when tested in comparison to other efficient algorithms and it appears to be one of the top heuristics for PBS. We ran tests to compare I_HSA with SGA which show that I_HSA always produces a schedule better than SGA. I_HSA's approximation ratio is,for some values of $d$ up to 10% better than the one of SGA. As in paragraph 5, we used five hundred test cases following the uniform distribution for a 30 source-30 destination system for values of setup cost varying from 0 to 100 and message durations varying from 0 to 120.

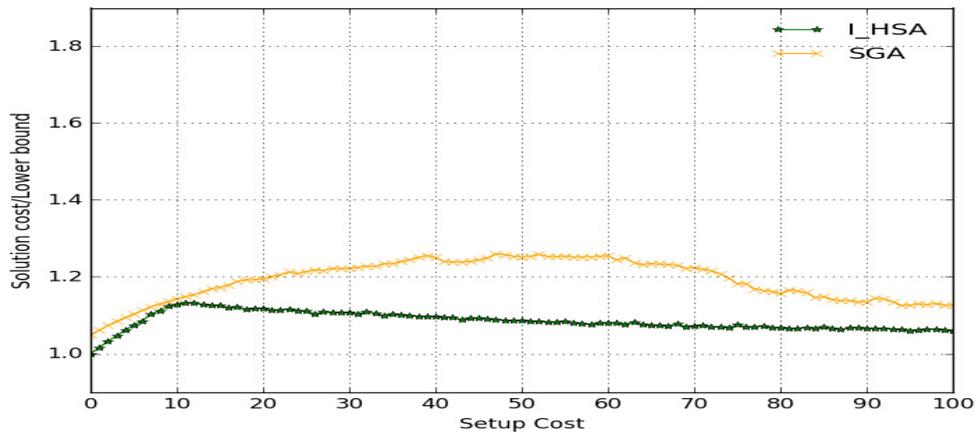

Figure 5: Comparison of I_HSA with SGA

# 9. CONCLUSIONS AND FUTURE WORK

Based on a reduction of a network transmission problem (PBS) to a scheduling problem (Open Shop) done by the authors in [2], we have designed a hybrid algorithm (I_HSA) using suitable variations of polynomial time algorithms for special instances of Open Shop (POSA and I_OS01PT) designed and critically improved for the purposes of this paper, in order to develop an efficient switch reconfiguration strategy for Multistage Interconnection Network transmissions. We have run tests to establish the efficiency of our hybrid algorithm and to suggest the appropriate value of network delay to switch from POSA to I_OS01PT. Knowing this value of d improves the computational complexity of I_HSA, which we fully illustrated. In these experiments we used datasets following uniform, normal and exponential distribution. In addition, we provided statistical data which gives additional insight on the performance of the variations.Furthermore, we tested I_HSA against SGA, one of the most efficient algorithms designed for PBS in the past to conclude that I_HSA's results have in most cases an approximation ratio up to 10% better than SGA's.

Future research could focus on further improvement of the time complexity of I_HSA. The fact that I_HSA's approximation ratio even for the worst data tested has always been less than 3/2, suggests that a formal mathematical proof of an approximation ratio lower than 2 might be possible.To further improve performance and complexity a new hybrid algorithm could be designed using different approaches on how to minimize each criterion separately.This algorithm might also be independent of the open shop approach. It would aim in minimizing just one of the criteria under the constraint that the other one is minimum.

## Authors

Stavros Birmpilis was born in Athens, Greece in 1994. Currently, he is a senior student in the School of Electrical and Computer Engineering at the National and Technical University of Athens, expecting to receive his diploma by July 2017. His research interests lie in the field of Algorithms and Discrete Mathematics.

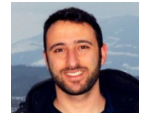

Timotheos Aslanidis was born in Athens, Greece in 1974. He received his Mathematics degree from the University of Athens in 1997 and a master's degree in computer science in 2001. He is currently doing research at the National and Technical University of Athens in the School of Electrical and Computer Engineering. His research interests comprise but are not limited to computer theory, number theory, network algorithms and data mining algorithms.

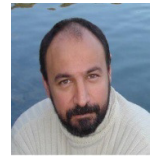